# The edge delamination of monolayer transition metal dichalcogenides

*Thuc Hue Ly,[†, ‡,]\* Seok Joon Yun,[‡] Quoc Huy Thi,[‡] and Jiong Zhao[†,‡,]\**

[†] Department of Applied Physics, The Hong Kong Polytechnic University, Hong Kong, China.

[‡] IBS Center for Integrated Nanostructure Physics, Institute for Basic Science, Sungkyunkwan University, Suwon 440-746, Korea.

[*]Corresponding author E-mail: (T. H. Ly) thuchue@skku.edu, (J. Zhao) zhaojiong@gmail.com

**Abstract**

Delamination of thin films from the supportive substrates is critical issues in thin film industry and technology. The emergent two-dimensional materials, atomic layered materials, such as transition metal dichalcogenides are highly flexible thus the buckles and wrinkles can be easily generated and play vital effects on the physical properties. Here we introduce one kind of patterned buckling behavior caused by the delamination from substrate initiated at the edges of the chemical vapor deposition synthesized monolayer transition metal dichalcogenides, mainly due to the thermal expansion mismatch. The atomic force microscopy and optical characterizations clearly showed the puckered structures associated with strains, whereas the transmission electron microscopy revealed the special sawtooth shaped edge structures which break the geometrical symmetry of the buckling behavior of hexagonal samples. The condition of this edge delamination is in accordance with the fracture theory. This edge delamination process and buckling upon synthesis is universal for most of the ultrathin two dimensional materials, and it is definitely noteworthy in their future applications.

**Keywords:** Tungsten disulfide, Molybdenum disulfide, edge delaminate, AFM, Tem, PL.

The emergent two-dimensional (2D) materials[1-3] consists of only single or a few atomic layers in thickness. Despite the well-established models of 2D materials are atomically flat[4], some experiments have also found noticeable out-of-plane fluctuations, especially in free-standing 2D materials[5,6]. Basically these 3D fluctuations which break the long range order should be responsible for their unexpected ultrahigh stability[7]. In another aspect, for those which are well supported by the rigid substrates, 2D materials are expected to fully comply with the surface morphology of the underneath substrates due to intrinsic flexibility[8,9]. And owing to this flexibility, wrinkles or buckles in 2D materials can be readily induced[8-11]. Herein we intend to address the edge effect on the buckling and wrinkling of 2D materials, more specifically, in the transition metal dichalcogenides (TMDs). In addition, our experimental characterizations of the monolayer TMDs and analysis clarified the previous substantial misunderstandings in "domain structures"[12] or "edge enhancement"[13] associated with their optical or chemical properties. We confirmed that the origin of the domain contrasts is probably not the atomic defects[12,13], rather the geometrical buckling in the flakes.

According to the thin film technology, the interfaces between the deposited films and the substrates occasionally have large strains, including lattice mismatch strain[14], thermal mismatch strain[15] or some other external loading induced strains[16]. The lattice mismatch strain often occurs in the metallic or covalent bonding interfaces whereas for van der Waals (vdW) 2D materials the thermal mismatch strain or external loading strain are prevalent. All the chemical vapor deposition (CVD) synthesized 2D TMD specimens need to be cooled down from high temperature to room temperature as the completion of synthesis[17,18], plus the 2D materials normally have distinct in-plane thermal expansion coefficients (CTE) compared to the substrates[19], considerable thermal mismatch strains can be expected between the 2D materials and the substrates. Therefore, the 2D materials we are studying are not the "as-grown" samples, but actually the post-growth strained samples.

Our $WS_2$ monolayers were grown on $Si/SiO_2$ substrates via CVD methods[20], specific growth details can be found in the experimental section. The optical microscopy (OM) (Supplementary Information Figure S1) and scanning electron microscopy (SEM) demonstrate the morphologies of the $WS_2$ flakes (Figure 1). Similar to all the previous literatures[21-24], and comparable with the other TMD family ($MoS_2$, $MoSe_2$, $WSe_2$), the monolayer $WS_2$ has well-faceted edges which preserves the three-fold symmetry of the crystal structure (point group: $D_{3h}$). Some of them are equatorial triangles, while others have truncated triangle shapes or even close to equatorial hexagons. Some previous reports have already discussed about the shape evolution for these TMD monolayers during growth[25], and we have prepared another manuscript on the growth mechanisms, hence we focus on the post-growth mechanical responses in this study.

Interestingly, the SEM secondary electron (SE) images of the $WS_2$ hexagonal flakes show shallow contrast differences between the nearby domains, even after one time Polymethylmethacrylate (PMMA) transfer process[26] to the new Si substrate (Figure 1c). The three-fold symmetry of the contrast matches well with the crystal symmetry, regardless of the six-fold symmetry in the hexagonal shape. In addition, the triangle $WS_2$ flakes do have similar contrast inhomogeneity within single flakes (Supplementary Information Figure S2). The SEM SE imaging is sensitive with the sample surface and charges but insensitive with atomic defects or chemical compositions[27], implying the non-uniform morphology in the hexagonal $WS_2$ flakes. As the next step, the as-grown samples were checked by atomic force microscopy (AFM) experiments. Figure 2a, b shows the AFM topographic images in contact mode (AFM-CM) and tapping mode (AFM-TM), respectively. The mono-layer $WS_2$ has thickness around 0.8 nm (Figure 2a). The topographic image indicates the as-grown $WS_2$ flake is flat when the AFM tip is forced to push the sample (in Contact Mode) (Figure 2a), however the height contrast in the domains emerges by the non-contact mode (Tapping Mode) (Figure 2b). This is the direct evidence of flake buckling in the brighter domains in Figure 2b. The friction force microscopy (FFM) of the as-grown $WS_2$ hexagonal flake is shown in Figure 2c. Unsurprisingly, The FFM image shows similar domain contrast as the SEM SE and AFM-TM images in the hexagonal flakes.

Note the edges of the hexagons can be divided into two types (Figure 1a), W-zigzag (W-ZZ) or S-zigzag (S-ZZ)[20]. The edge types of the domains can be determined by the transmission electron microscopy (TEM), two methods including selected area diffraction pattern (SAED) and high resolution scanning transmission electron microscopy (STEM) were employed (refer to Supplementary Information Figure S3, S4). The TEM results confirmed the dark (lower) contrast domains in SEM SE micrographs (Figure 1c), buckled domains in AFM-TM (Figure 2b) and high friction domains in FFM (Figure 2c) all correspond to the W-ZZ edge domains.

Further, high spatial resolution photoluminescence (PL) mapping on the $WS_2$ flakes (Figure 3) again exhibit the domain contrast as in the SEM and AFM results. There are distinct contrast in the full range integrated PL intensity for both as-grown sample (Figure 3a,b) as well as after-transfer samples to new Si wafers (Figure 3e). The W-ZZ edge domains all have much higher PL intensities than S-ZZ domains. Moreover, the A exciton peaks in the W-ZZ edge domains have ~2 nm blue shift compared to S-ZZ edge domains in the original sample (Figure 3c), in contrast a red shift of the A exciton peaks was observed in after-transfer sample (Figure 3f). The inset mapping images on the PL peak positions (Figure 3a,b,e) clearly revealed this opposite trend of peak shift for the W-ZZ and S-ZZ domains.

In a recent publication[12] the PL contrast in different domains was explained by the different atomic defect types. But the entire flake experience the same growth history, it is unlikely that during growth the diffusion of defects across the domain boundaries can be prohibited and abrupt boundaries in defect distributions are formed or even lead to opposite defect types (cation/anion defects). In our view, the enhancement of PL in the W-ZZ edges can be more reasonably explained by the buckling of W-ZZ domains as the AFM results, thus the suspended flakes can have higher PL abilities than the substrate supported ones[28]. Moreover, the PL intensity is remarkably enhanced on the edge of $WS_2$ (Figure 3a,b,e and Supplementary Information Figure S5), which was attributed to the higher defect concentration at edges previously[13]. The flake edge delamination from substrate can reproduce such PL behavior as well. Similar domain contrast including the edge effects are universally discovered in a lot of TMD materials such as $MoS_2$, $WSe_2$, etc[29,30]. It is also noted that the contrast between domains cannot be observed when the sample size getting smaller (<30 µm) (Figure 3d), especially for after-transfer samples. It is noteworthy after transfer process (both on silicon substrate), the PL intensity decreased almost 3 times and the A exciton peaks has blue shift for both W-ZZ and S-ZZ domains, implying the release of tensile strains during transfer.

To know better about the puckered structure, we employed TEM to examine the $WS_2$ monolayers which were transferred onto Quantafoil$^{TM}$ TEM grid via PMMA method[26]. Figure 4 presents the TEM images for the monolayer hexagonal $WS_2$ sample both in low magnification and medium magnification. The results show the W-ZZ edges are straight and atomically smooth (Figure 4d and Supplementary Information Figure S6), while the S-ZZ edges become sawtooth-like in mesoscale with the teeth period 50~200 nm (Figure 4c). The edge morphologies are mainly resulted from the difference in surface energies. The W-ZZ edges are more stable with lower formation energy, and the S-ZZ edges preferred to be decomposed into W-ZZ edge sections, forming the 60° sawtooth edges (Figure 4c). The high resolution annular dark field (ADF) images and diffraction patterns are utilized to analyze the crystal directions therefore the S-ZZ edges and W-ZZ edges are exclusively determined (Supplementary Information Figure S3, S4).

In addition, the dark field (DF) TEM images shows there are a lot of wrinkles on the $WS_2$ samples. DF TEM technique exhibit diffraction contrast hence very sensitive with the crystal defects or crystal tilting. They clearly demonstrate the wrinkles in monolayer $WS_2$. We name the wrinkles parallel to edge as "transverse wrinkles" (Figure 5a-d) while the wrinkles perpendicular to edge as "vertical wrinkles" (Figure 5e-h). We find the wrinkles mainly distribute close to the edges (Figure 5). Furthermore, both transverse and vertical wrinkles can occur near the W-ZZ edges, with vertical wrinkles in domination (Figure 5a-c), while only transverse wrinkles emerge near the S-ZZ sawtooth edges (Figure 5e-g). Note that all the above wrinkles are from the carbon film supported part, as the wrinkles can be significantly released in the suspended part (Supplementary Information Figure S7).

There are two mechanical processes involved in our experiments. One is the sample shrinking during cooling down in the CVD furnace, where the CTE mismatch between $WS_2$ monolayer and substrate cause significant interfacial strains in lateral direction. The second is the transfer from the growth substrate on to TEM grid, where the PMMA deposition changes the morphology of $WS_2$ flakes by vertical compression. Although the strain remained in the TMD flakes after growth have been notified by optical approaches[29], the origin and strain mechanisms for the buckles or wrinkles are yet to be elucidated. In overall, the material, growth temperature, size and shape of the flakes comprehensively influence the strain distributions after the first process.

It is widely accepted that the CTEs of 2D materials show much new features, such as graphene has a negative CTE[31,32], while the measured linear CTE from Raman spectroscopy of monolayer $WS_2$ around $10.3 \times 10^{-6}$ K$^{-1}$ [19] while the CTE of Si substrate is only $3.6 \times 10^{-6}$ K$^{-1}$ [33]. Therefore, we can expect a 0.4% tensile strain remained in the $WS_2$ flakes (Figure 6a) if assuming the cooling process is from 750 °C to room temperature. Actually the Raman mapping results on the $WS_2$ sample do show the up-shifted of $E_{2g}$ peaks in the flat parts (S-ZZ edges) (Supplementary Information Figure S8), caused by the in-plane tensile strain not yet released by delamination.

Both the $WS_2$ and Si have higher than threefold symmetry in basal plane ($WS_2$) or (111) surface (Si), so the thermal expansion and the residual elastic strain at the interface can be reckoned as isotropic[34]. And Si wafer is much thicker than $WS_2$ monolayer, hence strain in Si is negligible. If we assume the the $WS_2$ monolayer and Si is in ideal contact all the time, with no local displacement occurs at interface, the affine elastic deformation field for $WS_2$ should be homogeneous. In addition, for the 2D TMD materials family, the anharmonicity of $MoS_2$ and $WS_2$ are quite close[35], leading to similar magnitude of thermal expansion mismatch. However, due to the different temperature used to synthesize the TMD monolayers ($WS_2$ at 935 °C [20] > $MoS_2$ at 750 °C [36], after cooling the tensile strain remained in $WS_2$ flake is relatively larger than $MoS_2$.

The shape and size of the flakes are important because they are correlated with the delamination process of the flakes from the substrate. Delamination and buckling of the thin films with respect to the substrates under tensile stress ($\sigma$)[37] can be understood by existed theories[38,39]. Delamination or buckling are initialized at the edges (Figure 6). The simplest model to describe this is to consider a mode II cracking model under shear (Figure 6b). The substrate exerts shear stress on the edges of TMD monolayer which induce propagation of cracks. The critical crack size (*c*) along the edge has the relationship with the interfacial toughness ($K_{ic}$) between the film and substrate[40]:

$$c = \frac{K_{ic}}{\pi \sigma^2}, \qquad (\text{eq. 1})$$

with initial crack size larger than *c* along the edge, it is able to catastrophically propagate toward the inner part of the flake and then lead to buckling. And if the crack starts to propagate, the crack can continue in the radial direction (Figure 6c) because of the ease of tensile strain release perpendicular to the free edge, whereas in the circumferential direction the tensile strain cannot be released therefore the crack will be ceased if the crack line gradually turns into the radial direction (Figure 6c). The straight W-ZZ edges (Figure 4d) can have larger initial cracks which is over *c* compared to the sawtooth-like S-ZZ edge (Figure 4c), as on the sawtooth edge the initial crack size are interrupted by the periodical sawtooth shape, with the period (~50 nm) as the upper limit of initial crack size. This explains why the delamination always occur first in the straight W-ZZ edges, and tend to follow threefold symmetry for the whole buckling process.

The size of the flakes also matters as larger flakes can have higher possibility to have larger initial crack and then trigger the delamination. On the contrary, small flakes without sufficiently large initial crack with the size over *c* usually have less buckles. Furthermore, in $WS_2$ it is easier to find such domain patterns than in $MoS_2$ (Supplementary Information Figure S9), in agreement with the greater thermal expansion mismatch strain ($\sigma$) at the interface and smaller critical crack size (*c*) for $WS_2$ delamination.

The lower SEM SE contrast for the buckled parts (Figure 1) may be due to the reduction of charge scattering at the interface between suspended 2D materials and substrates. The domain contrast in the AFM-TM image and FFM image (Figure 2) can be straightforwardly explained by the buckles in the W-ZZ domain and subsequent larger friction force on the AFM tip (Figure 2c), similar to the previous analysis of friction origin on graphene[41]. The buckles alter the local PL properties such as PL intensity and peak position[42]. In our case, the difference of PL intensity between non-buckled part and buckled part can be rationalized by the enhanced PL ability for suspended TMD monolayers than the supported ones[27], which comes from the reduced doping effect in suspended membranes. The PL peaks blue shift in the after-transfer samples compared to without-transfer samples (Figure 3) is in agreement with the release of significant thermal mismatch tensile strains.

In the second mechanical process, the deposition of PMMA upon the buckling parts in $WS_2$ will introduce wrinkles as shown in Figure 5. In the W-ZZ buckled domains, the wrinkles parallel as well as perpendicular to the edges can be formed. However, in the flat S-ZZ domains, vertical wrinkles can be precluded because of the restriction of circumferential deformation, however the free relaxation of the edges in radial direction can naturally form the transverse wrinkles. It should be noted that before transfer of $WS_2$ onto the TEM grid, the changes from buckles into wrinkles may already occur in some samples. And the buckles before transfer sometimes can be fully kept even after the transfer process[12]. Some factors such as sample quality and conditions of transfer matters. Moreover, the environmental attack on the $WS_2$ samples such as light and humid can introduce defects[43, 44] and significantly reduce the strain and flexibility of the flakes, hence suppressing the delamination and puckering. The degree of delamination is also dictated by the interaction between substrate and 2D membrane, i.e., on the elastomer substrate, the TMD layers can be continuously elongated until 16% tensile strain without relaxations[45,46].

In summary, there are large quantities of open questions in the inhomogeneity of strain distribution and related physical properties in 2D materials. In this work we have addressed the importance of the edge delamination and the subsequent puckering such as buckles and wrinkles in single atomic layers. The "domain contrast" and "edge enhancement" in 2D materials should involve considerations on the buckling patterns in special geometries. The atomic defect which was the main concern previously may not be the dominant reason. They may only act as the consequences of buckles, because the flat and buckled parts have different chemical activities, inducing defects under ambient conditions. After all, herein we have addressed the delamination and buckling issues which can play a vital role in the synthesis as well as in the optical, electrical or chemical applications for these emergent 2D materials.

*Experimental Section:*

**Synthesis of tungsten disulphide ($WS_2$) on $SiO_2$/Si wafer**. $WS_2$ was grown on $SiO_2$/Si wafer by atmospheric CVD process. For synthesizing monolayer $WS_2$, we first coated precursor solution on $SiO_2$/Si substrate by spin casting method. Preparation of precursor solution was conducted by mixing three type of water based solution. (three type of solution are defined as A, B and C), whereas, A (Tungsten precursor) includes 0.1 gram of Ammonium metatungstate hydrate [$(NH_4)_6H_2W_{12}O_{40}\cdot xH_2O$ : Sigma-Aldrich, 463922] dissolved in 10ml of DI water, B (promoter) contains Sodium cholate hydrate (Sigma-Aldrich, C6445) as a promoter was dissolved in DI water (0.3 g of SC in 10 ml of DI water) and C (Medium solution) is a medium to mix promoter and precursor which was prepared from OptiPrep density gradient medium (Sigma-Aldrich, D1556, 60 % (w/v) solution of iodixanol in water). C does not affect growth but only for better spin casting process. A, B and C solutions were mixed in the certain ratio for its purpose (discussed below). Then, the mixed solution was coated onto $SiO_2$/Si wafer by spin-casting at 3000 rpm for 1 min.

A two-zone CVD system was introduced for controlling sulfur and substrate zone temperature separately. Here, 0.2 gram of sulfur (Sigma, 344621) was loaded, while the solution coated substrate containing metal precursor was placed in another zone. Synthesis of $WS_2$ in this work was carried out at atmospheric pressure. For growth, sulfur zone was heated up to 210 °C at a rate of 50 °C/min at the same time, the substrate zone was set to 780 °C. 600 sccm of Nitrogen and 5~20 sccm of Hydrogen gas were introduced as carrier gas and reactive agent.

**Shape controlling of CVD grown $WS_2$ on $SiO_2$/Si wafer.** For synthesizing triangular shape of $WS_2$, precursor solution at a ratio of 1:6:1 was spin-casted on $SiO_2$/Si wafer. When substrate temperature reached to maximum (800 °C), 5sccm of hydrogen was introduced for 10 minutes.

The key factor to control shape of $WS_2$ is hydrogen injection timing. Hydrogen accelerate growth rate of $WS_2$ by enhancing reduction rate of tungsten oxide. It also effects on etching process. For hexagon shape of $WS_2$, Hexagonal shape $WS_2$ case, precursor solution ratio is set to be 2:6:1 and 10 sccm of hydrogen was introduced from the beginning.

**Photoluminescent (PL) and Raman spectroscopy**. PL and Raman mapping (NT-MDT, 532 nm wavelength, NTEGRA Spectra PNL, x100 lens, 0.7 N.A.) was performed using a laser (532 nm) with ~ 30 µW power. The scanned image was obtained at 128 x 128 pixels with a grating of 1800 g/mm to yield a spatial resolution of 200 nm for confocal PL and 600 g/mm to yield a spectral resolution of < 0.1 $cm^{-1}$ for confocal Raman mapping, respectively. The accumulation time for each spectrum was 0.3 second for image scanning and 0.5 seconds for a single spectrum. An area filter was used to extract PL intensity map (580 to 680 nm) and Raman spectrum map with an integration of $E^1_{2g}$ peak (335 to 380 $cm^{-1}$), $A_{1g}$ peak (405 to 430 $cm^{-1}$).

**Atomic force microscopy (AFM)**. AFM images were obtained using a SPA400 system (SEIKO, Japan) in tapping mode for observation of only topography image and in contact mode for coherent observation of topography image and friction force image. A Au tip (MikroMasch, Estonia) with an approximately 10 nm tip radius was used. The force constant and resonant frequencies of the tips were approximately 1.6N/m and 28 kHz. To get the friction image, a constant force ~ -1nN was applied.

**Scanning electron microscopy (SEM).** Field-emission scanning electron microscopy (FESEM) (JSM7000F, Jeol, Japan) was used to examine the surface morphology of samples at different accelerating voltages to obtain a high level of contrast at different magnifications. An accelerating voltage of 10 keV was used to obtain sufficiently pronounced signals while retaining sensitivity to the sample surface.

**TEM sample preparation.** The CVD tungsten disulphide was transferred on a hole with 1.2 µm in diameter in Cu quantifoil TEM grid (Product No. 658-200-CU) by PMMA-assistant method. Thin layer PMMA was spin-coated on as-grown $WS_2$/SiO2/Si substrate (2000 rpm, 1min). The $WS_2$ and PMMA support were then detached from the $SiO_2$/Si substrate by floating the PMMA/$WS_2$/$SiO_2$/Si, with the PMMA side up, in a 1M HF solution. Next, PMMA/$WS_2$ was washed by deionized water. The PMMA/$WS_2$ layer is cooped out in pieces onto TEM grid, then

PMMA was removed gently by evaporated acetone (acetone was heated up to 130 ºC), leaving $WS_2$ suspended freely on holes in TEM grid substrate. Finally, sample was annealing at 180 ºC in a high vacuum ($10^{-6}$ Torr) during 12 hours to further remove PMMA.

**TEM measurement.** TEM experiments were carried out using a JEM ARM 200F machine under 80 kV. The acquisition time for dark field (DF) imaging was 1 s using the smallest objective lens aperture. And the reflex (10-10) was always selected for DF imaging. The HR-TEM imaging acquisition time was also 1 s. ADF-STEM. Annular dark field (ADF)-STEM imaging was conducted with a CEOS aberration-corrector on the same TEM. High-angle annular dark field (HAADF) images were acquired at a 20 mrad convergence angle

ACKNOWLEDGMENT

This work was supported by the Hong Kong Polytechnic University Grant (No. G-YW1U, No. 1-ZE8C) and Institute for Basic Science (IBS-R011-D1).

**FIGURE LEGENDS:**

**Figure 1. The schematic and SEM characterizations of the WS$_2$ flakes on SiO$_2$ substrates.** (a) The cartoon of the hexagonal buckled monolayer WS$_2$. The in-plane tension is highlighted by arrows. Two insets show the atomic structures of two different type edges, non-buckled S-ZZ edge and buckled W-ZZ edge. (b) The secondary electron images of the as-synthesized WS$_2$ flakes. Scale bar is 20 µm. (c) Higher magnification secondary electron images for the "six patch" domain contrast in hexagonal WS$_2$ monolayer after transfer to new silicon substrate. The white dot in the center is marker on silicon wafer. Scale bar is 10 µm.

**Figure 2**. **AFM characterizations of the WS$_2$ monolayers.** (a) The contact mode (CM) topographic AFM image of as-synthesized monolayer WS$_2$. The inset shows the height profile across the edge. (b) The tapping mode (TM) topographic AFM image of monolayer WS$_2$ with domain contrast after transferred. (c) The friction force microscopy (FFM) image of as-synthesized monolayer WS$_2$ with domain contrast. All scale bars are 10 µm.

**Figure 3**. **The PL characterizations of monolayer WS$_2$ flakes.** (a,b) The PL intensity mapping image for the as-synthesized hexagonal and quasi-triangle monolayer WS$_2$ flakes. Inset shows the corresponding PL peak position mapping, with wavelength scale bars on the right side. Scale bar of (a) is 7 µm, and (b) is 12 µm. (c) The corresponding PL spectra for the W-ZZ edge domain and S-ZZ edge domain in (a). (d,e) The PL intensity mapping image for two WS$_2$ flakes which are transferred to new silicon substrates. Inset shows the corresponding PL spectra peak position mapping, wavelength scale bars on the right side. scale bar in (d) is 6 µm, (e) is 12 µm. (f) The PL spectrums for the transferred WS$_2$ sample in different domains in (e).

**Figure 4**. **TEM characterizations of WS$_2$ flakes.** (a) Low magnification TEM image for the WS$_2$ monolayer sample. Scale bar is 30 µm. The dashed circles highlight the zones which are magnified by the dark field images in (b-d). (b-d), The dark field images for the different types of edges, the dashed lines highlight the sawtooth edge and straight edge. scale bars are 500 nm.

**Figure 5**. **The wrinkles in WS$_2$ monolayers characterized by TEM.** (a,b) TEM dark field images for the W-ZZ edges, the white triangles mark some vertical wrinkles close to edge. (c) Dark field images of inner part but still in the W-ZZ edge domain, with white arrow shows the W-

ZZ edge direction and white triangles mark some vertical wrinkles. (d) Scheme of the vertical wrinkles on the W-ZZ edge. (e,f) TEM dark field images for the S-ZZ edges, the white triangles mark some transverse wrinkles close to edge. (g) Dark field images of inner part but still in the S-ZZ edge domain, with white arrow shows the S-ZZ edge direction and white triangles mark some transverse wrinkles. (h) Scheme of the transverse wrinkles on the S-ZZ edge. All scale bars are 200 nm.

**Figure 6**. **Schematic of delamination process initialized from the edge.** (a) The initial crack length over the critical crack size at the edge will start to propagate toward inner part of the flake. (b) Cross-section view of the mode II crack, with black arrow showing the crack propagation direction. (c) The develop path of the crack lines follows the numbered dashed black lines in the order 1,2,3, until forming the quasi-triangle shape delaminated and buckled domains.
.

**Figures:**

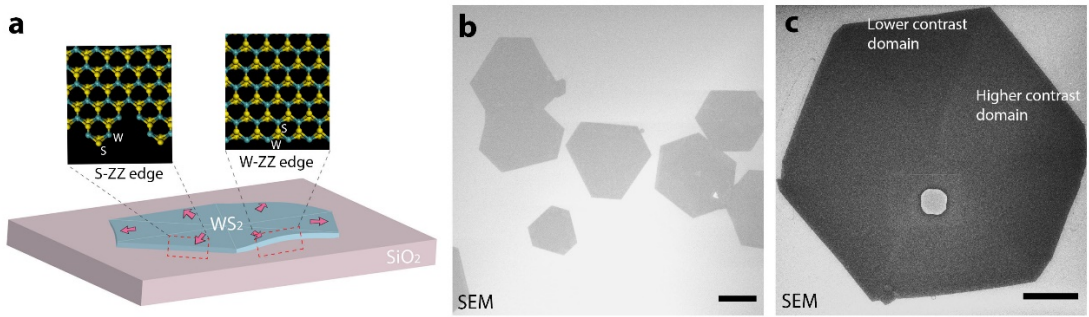

**Figure 1**

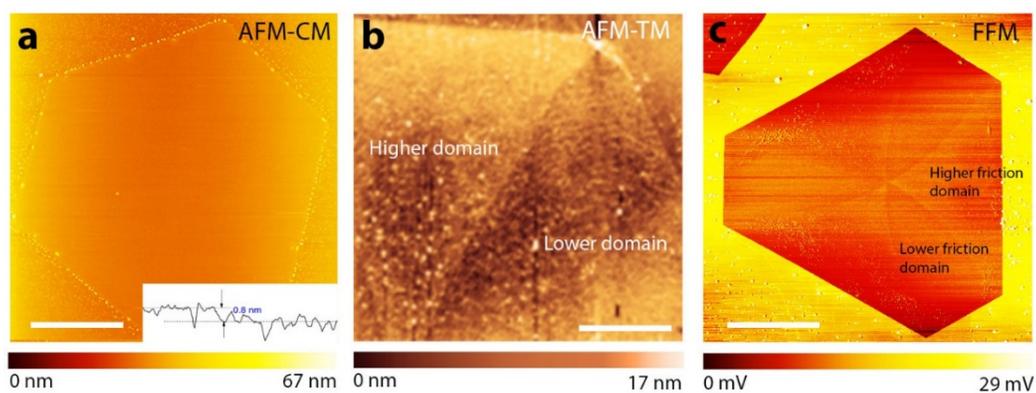

**Figure 2**

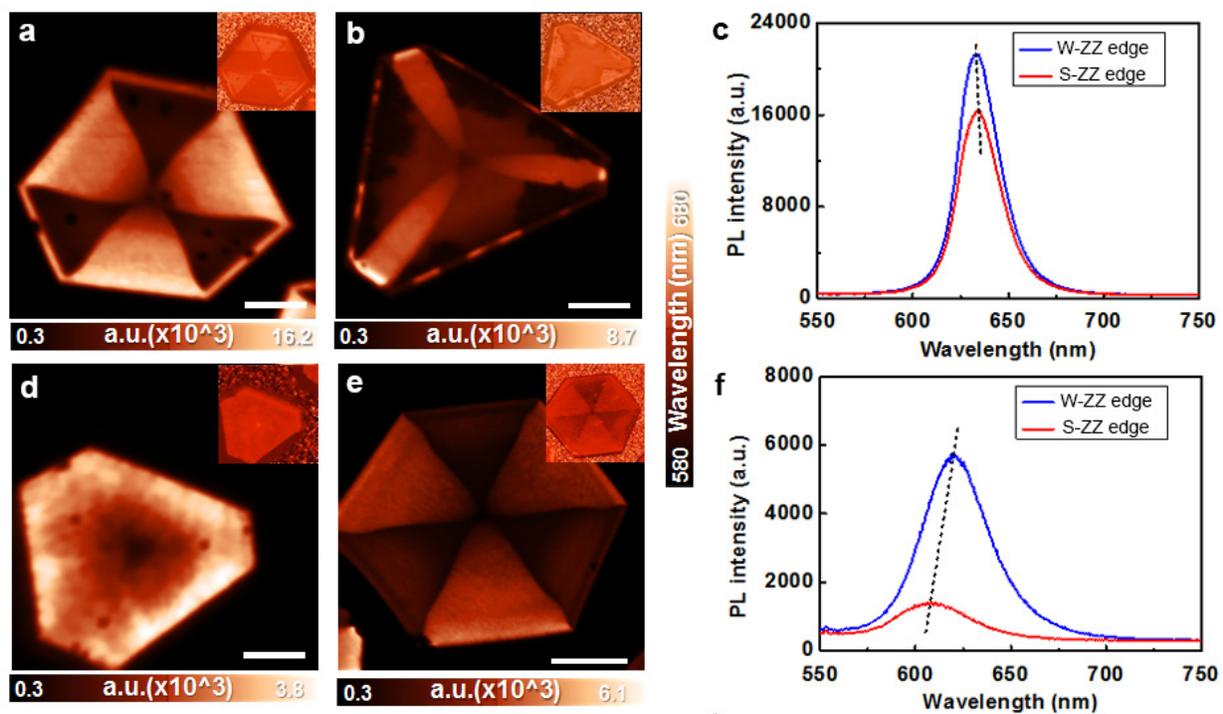

**Figure 3**

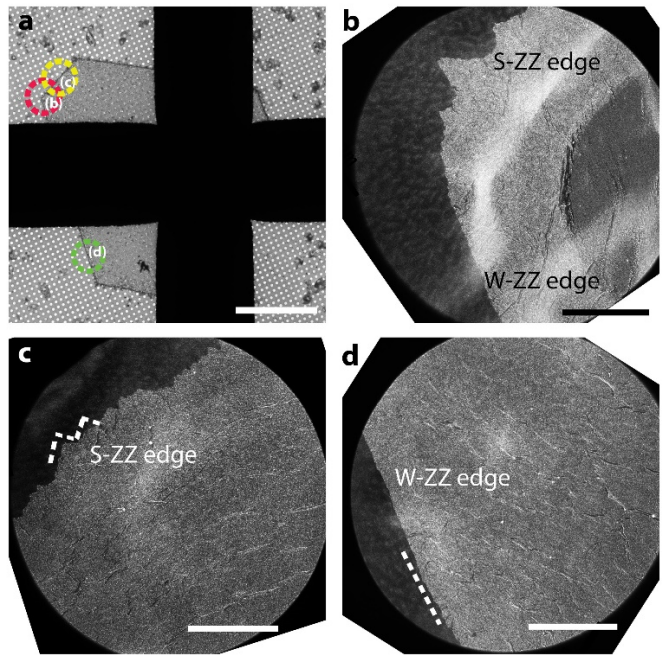

**Figure 4**

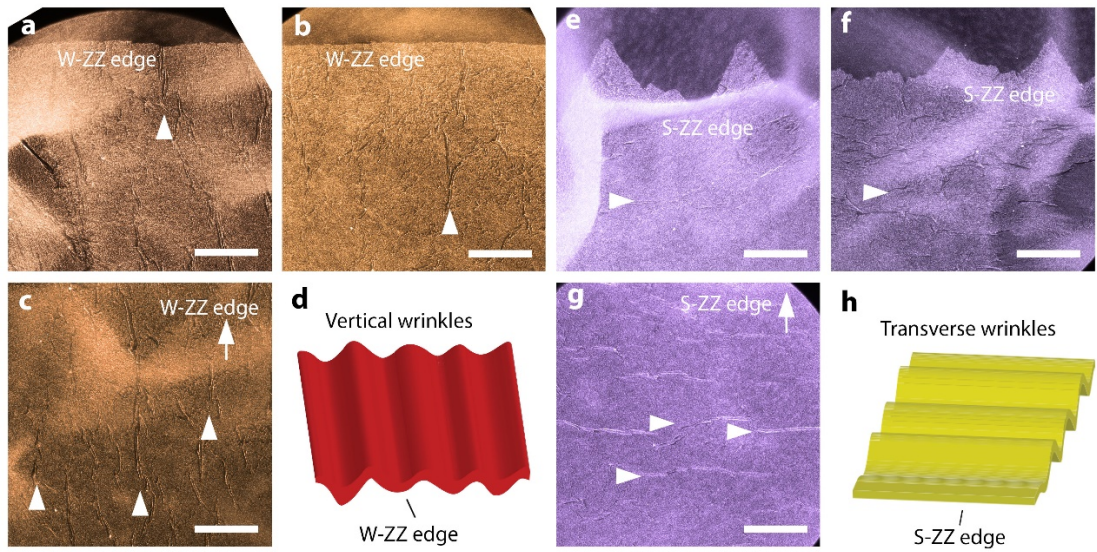

**Figure 5**

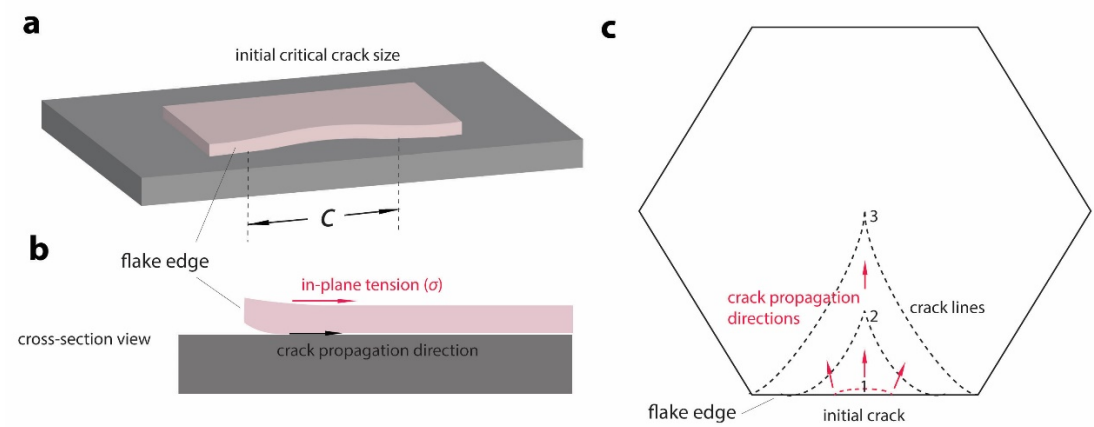

**Figure 6**